\documentclass[conference]{IEEEtran}
\IEEEoverridecommandlockouts
\usepackage[explicit]{titlesec}
\usepackage{graphicx}
\usepackage{changes}
\usepackage{color, soul}
\usepackage{booktabs} 
\usepackage{array} 
\usepackage{svg}

\usepackage{epstopdf}
\usepackage{subfigure} 
\usepackage{algorithmic}
\usepackage{amsmath}
\usepackage{amsthm}
\usepackage{cite}

\usepackage{bbm}
\usepackage{hyperref}
\usepackage{amssymb}

\newcommand{\RNum}[1]{\uppercase\expandafter{\romannumeral #1\relax}}

\usepackage{stfloats}

\usepackage{color}
\usepackage{tikz,xcolor,hyperref}

\definecolor{lime}{HTML}{A6CE39}

\titlespacing{\section}{0pt}{1.2ex plus .0ex minus .0ex}{.3ex plus .0ex}
\titlespacing{\subsection}{0pt}{1.2ex plus .0ex minus .0ex}{.3ex plus .0ex}

\makeatletter
\DeclareRobustCommand{\orcidicon}{%
	\begin{tikzpicture}
	\draw[lime, fill=lime] (0,0) 
	circle [radius=0.16] 
	node[white] {{\fontfamily{qag}\selectfont \tiny ID}};    \draw[white, fill=white] (-0.0625,0.095) 
	circle [radius=0.007];    \end{tikzpicture}
	\hspace{-2mm}}
\foreach \x in {A, ..., Z}{%
\expandafter\xdef\csname orcid\x\endcsname{\noexpand\href{https://orcid.org/\csname orcidauthor\x\endcsname}{\noexpand\orcidicon}}
}

\newcommand*\bigcdot{\mathpalette\bigcdot@{.5}}
\newcommand*\bigcdot@[2]{\mathbin{\vcenter{\hbox{\scalebox{#2}{$\m@th#1\bullet$}}}}}
\makeatother

\usepackage[linesnumbered,ruled,vlined]{algorithm2e}

\begin{document}
\title{Joint Transmission and Control in a Goal-oriented NOMA Network}

\author{ 
Kunpeng Liu$^1$, Shaohua Wu$^{1, 2}$, Aimin Li$^1$, and Qinyu Zhang$^{1, 2}$ \\
$^1$ Harbin Institute of Technology (Shenzhen), Shenzhen 518055, China \\
$^2$ Peng Cheng Laboratory, Shenzhen 518055, China \\
e-mail: 23S052017@stu.hit.edu.cn, liaimin@stu.hit.edu.cn, hitwush@hit.edu.cn, zqy@hit.edu.cn

	\vspace{-0.5em}
\thanks{
This work has been supported in part by the Guangdong Basic and Applied Basic Research Foundation under Grant no. 2022B1515120002, and in part by the National Natural Science Foundation of China under Grant nos. 61871147, 62027802, and in part by the Major Key Project of PCL under Grant no. PCL2024A01.}
}

\maketitle

\begin{abstract}
Goal-oriented communication shifts the focus from merely delivering timely information to maximizing decision-making effectiveness by prioritizing the transmission of high-value information. In this context, we introduce the Goal-oriented Tensor (GoT), a novel closed-loop metric designed to directly quantify the ultimate \textit{utility} in Goal-oriented systems, capturing how effectively the transmitted information meets the underlying application’s objectives. Leveraging the GoT, we model a Goal-oriented Non-Orthogonal Multiple Access (NOMA) network comprising multiple transmission-control loops. Operating under a pull-based framework, we formulate the joint optimization of transmission and control as a Partially Observable Markov Decision Process (POMDP), which we solve by deriving the belief state and training a Double-Dueling Deep Q-Network (D3QN). This framework enables adaptive decision-making for power allocation and control actions. Simulation results reveal a fundamental trade-off between transmission efficiency and control fidelity. Additionally, the superior utility of NOMA over Orthogonal Multiple Access (OMA) in multi-loop remote control scenarios is demonstrated.
\end{abstract} 


\IEEEpeerreviewmaketitle

\section{Introduction} 
In the forthcoming neXt-generation ultra-reliable low-latency communication (xURLLC) in 6G networks, information freshness has garnered significant research interest owing to its essential role in supporting mission-critical applications, including industrial automation, smart factories, and smart healthcare \cite{10173693}. In these xURLLC scenarios, wireless sensors regularly transmit updates on monitored phenomena to a control entity. The extensive deployment of edge nodes (e.g., sensors, controllers, actuators) across the network poses challenges in efficiently leveraging the sufficient yet still lacking communication resources.

To address these challenges, Non-orthogonal Multiple Access (NOMA) presents an effective solution by supporting a significantly higher number of users and offering greater frequency efficiency compared to traditional Orthogonal Multiple Access (OMA) systems. As an essential metric considered in 6G-xURLLC\cite{10173693}, the Age of Information (AoI) was initially proposed in \cite{6195689} and has subsequently garnered substantial research attention. Numerous research has concentrated on evaluating AoI performance in NOMA systems. \cite{8845254} is an early study to evaluate AoI performance between NOMA and OMA, which demonstrates that the spectral efficiency advantages of NOMA do not consistently lead to AoI performance improvements. Building on this insight, considerable research like \cite{9174163,10260312,10225923,9875026} has explored various techniques aimed at further enhancing AoI performance within NOMA systems. Authors from \cite{9174163} investigate a hybrid NOMA/OMA system and employ a Markov Decision Process (MDP) framework to enable adaptive switching between NOMA and OMA modes. \cite{10260312} derives a closed-form expression for AoI in NOMA assisted grant-free transmission, revealing the superior AoI performance of NOMA compared to OMA. \cite{10225923,9875026} enhance transmission reliability by incorporating hybrid automatic repeat request(HARQ), thereby improving AoI performance of the NOMA system.


In future 6G networks, the exponential increase in access nodes and the unprecedented surge in data volume exert immense pressure on traditional bit-driven paradigms. In response, \textit{Goal-oriented} communication, which prioritizes overall system utility and the ultimate goal achieved, has rapidly attracted widespread attention. AoI is also a key indicator for assessing utility; however, it has limitations in capturing dynamic changes in source states. To address this, researchers have introduced a range of extended metrics, including Age of Changed Information (AoCI) \cite{9565910}, Age of Synchronization (AoS) \cite{8437927}, and Age of Incorrect Information (AoII) \cite{9137714}, among others. These metrics provide a more comprehensive characterization of system utility from various perspectives. However, the aforementioned metrics are all based on end-to-end open-loop systems which overlook the actuation timeliness. Some novel metrics have been developed to address this limitation, the Age of Loop (AoL) is proposed in \cite{9569366}, which evolves based on both uplink state updates and downlink command. \cite{9475174} introduced the Cost of Actuation Error (CAE), which quantifies the cost incurred by the controller due to inaccurate real-time estimations. Although these metrics account for closed-loop systems, they do not consider how decisions are made at the receiver and how these decisions impact the source. To date, the relationship between these open-loop indicators and the utility of goal-oriented decision-making remains an open question.

To address this gap, our previous work \cite{wcmli} proposed the Goal-oriented Tensor (GoT),  which incorporates a context dimension, thereby facilitating a nuanced and flexible characterization of utility across a multitude of scenarios. Moreover, the GoT can seamlessly degenerate into established utility metrics, achieving a harmonious integration that underscores its versatility. Crucially, it comprehensively considers the impact of receiver decisions on the source, offering an in-depth evaluation of system utility within goal-oriented communication frameworks.

Motivated by the above, we focus on a goal-oriented example within a remote control scenario. Our objective is to design a transmission-control joint optimization scheme aimed at minimizing the average GoT in a Goal-oriented NOMA network. Specifically, this paper achieves three-fold contributions:
\begin{itemize}
    \item We propose the model for a Goal-oriented NOMA network and quantify system utility by characterizing the network's GoT.
    \item We implement transmission-control joint optimization by minimizing the long-term average GoT. Specifically, the sequential decision-making problem is formulated as a Partially Observable Markov Decision Process (POMDP), from which the belief state is further derive. Finally, a Double-Dueling Deep Q-Network (D3QN) is trained to enable intelligent power allocation and control decisions.
    \item We reveal the inherent trade-off between transmission and control in remote control system. Moreover, our results demonstrate that NOMA achieves superior system utility compared to OMA in remote control applications.
    
\end{itemize}

\section{System Model}\label{section II}
\begin{figure}
    \centering
    \includegraphics[angle=0,width=0.5\textwidth]{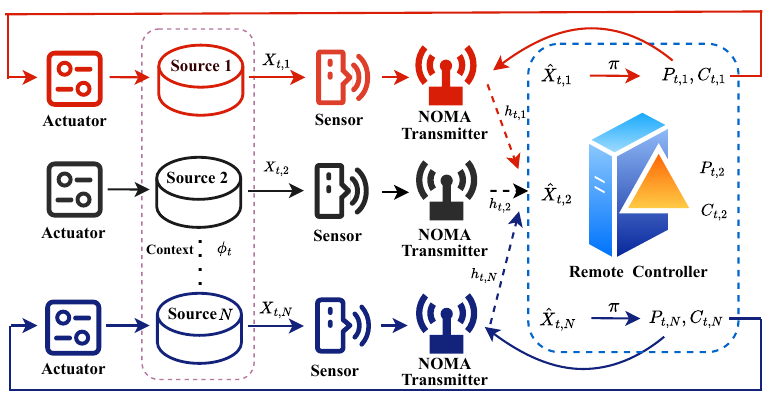}
    \caption{Goal-oriented NOMA Network Model.}
    \label{fig:1}
\end{figure}

As shown in Fig. \ref{fig:1}, we consider a time-slotted multi-loop transmission-control system with $N$ controlled monitored proceedings. Sensors are assumed to capture the semantic state of each source perfectly, denoted as $X_{t,i}$, where $i\in[N]$, within a given shared context $\phi_t$. At each time slot, the NOMA transmitters, following the \textit{pull-based} framework, transmit each corresponding sensor's signal with power $P_{t,i}$ over an unreliable channel, as specified by the remote controller. Upon receiving and decoding the superimposed signals, the remote controller allocates transmission power to $N$ transmitters and generates the corresponding control commands. These commands are then transmitted via the downlink\footnote{ We assume that the downlink transmission is error-free, a hypothesis also made in \cite{8329618,1273769}.}  to $N$ actuators. The actuators implement these commands, thereby influencing the sources' future evolution and enhancing the overall \textit{utility} of the system. We remark that the goal of this communication process is not merely accurate transmission but rather the facilitation of effective actions that enhance system \emph{utility}.

\subsection{Semantics and Context Dynamics}
We consider each semantic source as a controlled discrete Markov source, where $X_{t,i}\in \mathcal{X}=\left\{ x_1,x_2\cdots,x_{|\mathcal{X}|} \right\}$ represents the state of source $i$ at time slot $t$. The evolution of each source is influenced by the corresponding actuation $C_{t,i}$, where $C_{t,i}\in\mathcal{C}=\{c_1,c_2,\cdots,c_{|\mathcal{C}|}\}$. This dynamics is characterized by the following transition probability:
\begin{equation}
\Pr\left( X_{t+1,i}=x_n|X_{t,i}=x_m,C_{t,i} =c_k \right) =p_{m,n}^{(k)}.
\label{eq:1}
\end{equation} 
The context $\phi_t\in\mathcal{V}=\{ v_1,v_2\cdots,v_{|\mathcal{V}|}\}$ represents the environmental conditions surrounding the monitored sources. We assume that its evolution is independent of the actuation taken\cite{litcom}, and the dynamics of $\phi_t$ is characterized by a Markov chain with transition probability:
\begin{equation}
\Pr\left( \phi_{t+1}=v_b|\phi_t=v_a \right) =p_{a,b}.
\label{eq:1}
\end{equation} 

\subsection{NOMA Uplink and Error Rate}\label{sectionII}
Without loss of generality, the distances between the NOMA transmitters and the controller are assumed to satisfy $d_1\le d_2\le d_3\le\cdots\le d_N$. We consider a wireless fading channel where transmitted signals experience quasi-static Rayleigh fading. Accordingly, the channel coefficient from transmitter $u_i$ to the controller at time slot $t$ is represented as $h_{t,i}=\sqrt{d_i^{-\tau}}g$, where $g \sim \mathcal{C} \mathcal{N} \left( 0,1 \right)$ denotes a unit Rayleigh channel coefficient with zero mean and variance $\sigma^2 = 1$, and $\tau$ is the path loss exponent. In the NOMA uplink, multiple signals can be superimposed and transmitted over the same time-frequency resource block. We consider the transmission delay is one time slot, then the signal received by the controller is given by:
\begin{equation}
y_{t+1}=h_{t,i}\sum_{i=1}^N{\sqrt{P_{t,i}}\vartheta _{t,i}}+N_t,
\end{equation}
where $\vartheta _{t,i}$ represents the unit-power signal obtained after modulation of $X_{t,i}$, and $N_t\sim\mathcal{CN}(0,1)$ denotes the unit complex additive white Gaussian noise (AWGN).

The controller decodes the superimposed signals using successive interference cancellation (SIC). To ensure the decoding success rate, the controller begins by decoding the signal with the highest received power $P_{t,i}^{(rx)}=P_{t,i}|h_{t,i}|^2$, removes it from the composite signal, and then proceeds to decode the next strongest signal. This process repeats iteratively until all signals are successfully decoded. Without loss of generality, we assume $P_{t,1}^{(rx)}\ge P_{t,2}^{(rx)}\ge\cdots\ge P_{t,N}^{(rx)}$. In this case, the Signal-to-Interference-plus-Noise Ratio (SINR) for the transmitter $u_i$ can be expressed as:

\begin{equation}
\gamma _{t,i}=\frac{P_{t,i}^{(rx)}}{\sum\limits_{j=i+1}^N{P_{t,j}^{(rx)}+\sigma ^2}}.
\end{equation}

For many real-time applications, short channel codes play a crucial role in enabling ultra-reliable low-latency communication (URLLC) links. Accordingly, we consider channel coding within the finite block length (FBL) regime. In each SIC decoding round, the error probability for the controller decoding the signal from transmitter $u_i$ under a fixed code length $m$ can be approximated as \cite{polyanskiy2010channel}:
\begin{equation}
	\label{eq:7}
\tilde{\varepsilon}_{t,i}\approx Q\left( \frac{C\left( \gamma _{t,i} \right) -n/m}{\sqrt{V\left( \gamma _{t,i} \right) /m}} \right),
\end{equation}
where $n$ is the message length sent to $u_j$, $C( \gamma _{t,i})=\log _2(1+\gamma_{t,i})$ is the channel capacity, and $V(\gamma _{t,i})$ represents the channel dispersion, given as $V(\gamma _{t,i})=(1-1/(1+\gamma _{t,i})^2)(\log_2e)^2$.
According to the SIC, successful decoding of $u_i$'s signal depends on the sequentially error-free decoding of all higher-power signals. Consequently, the overall error probability for $u_i$ can be expressed as follows:
\begin{equation} \label{eq:10}
\varepsilon _{t,i}=1-\prod_{j=i}^N{( 1-\tilde{\varepsilon}_{t,j})}\,.
\end{equation}

\subsection{Goal-oriented Transmission and Control}\label{subsectionC}
In this paper, we consider a \textit{pull-based} communication framework in which NOMA transmission power scheduling is triggered by commands from a remote controller rather than proactive resource allocation by the transmitter. In this framework, the controller dynamically requests updates and allocates transmission power to sensor signals as needed, aiming to maximize control performance by selectively refreshing its information at strategic intervals. This pull-based approach enables joint design of both transmission and control. 

In the \textit{pull-based} framework, the controller does not have direct, real-time access to each sensor’s current semantics $X_{t,i}$; instead, it can only observe the most recently received state $\hat{X}_{t,i}={X_i(t-\varDelta_{t,i})}$, along with its associated age of information (AoI), denoted as $\varDelta_{t,i}$, and $\varDelta_{t,i}\in\mathcal{T}=\{1,2,\cdots,\varDelta_{\max}\}$.\footnote{A \textit{timestamp} is added to each transmitted packet, enabling the remote controller to directly determine the AoI based on the \textit{timestamp} information.}

We remark that the observation $(\hat{X}_{t,i},\varDelta_{t,i})\in\mathcal{X}\times\mathcal{T}$ is strongly dependent on transmission, which can be characterized as follows:
\begin{align}
	\label{eq:2}
\varDelta _{t,i} &=\begin{cases}
	\varDelta _{t,i} +1,& w.p.\,\, \varepsilon _{t,i}\\
	1, & w.p.\,\, 1-\varepsilon _{t,i}\\
\end{cases}\,,\\
	\label{eq:3}
\hat{X}_{t+1,i}&=\begin{cases}
	\hat{X}_{t,i},& w.p.\,\, \varepsilon _{t,i}\\
	X_{t,i}, & w.p.\,\, 1-\varepsilon _{t,i}\\
\end{cases}\,.
\end{align}

At each time slot $t$, the controller utilizes the \textit{observation history} $\mathcal{H}_{o,t}(\hat{X}_{t',i},\varDelta_{t',i})_{t'\le t,i\in[N]}$ and the \textit{action history} $\mathcal{H}_{a,t}=(C_{t',i},P_{t',i})_{t'\le t-1,i\in[N]}$ to generate both transmission power scheduling $P_{t,i}$ and controll action $C_{t,i}$. Higher-priority or more urgent information is allocated higher transmission power to ensure timely updates. The controll action $C_{t,i}$ is subsequently executed by the actuator, influencing the evolution of the semantic source and the system \emph{utility}.

\section{Problem Formulation}\label{sectionIII}
\subsection{Metric to Characterize the Goal}
The goal of the system is to ensure the semantic sources remain in the optimal state corresponding to the context $\phi_t$ for as long as possible, while minimizing both power expenditure and control costs. To evaluate the multiple closed-loop \textit{utility} that considers the impact of decision-making on the source, we utilize the Goal-oriented Tensor (GoT) as a metric, which comprises the following components \cite{wcmli,litcom}:  
\begin{itemize}
    \item{\textbf{Status Cost $C_1(X_{t,i},\phi_t)$}}. It represents the intrinsic cost of semantic status $X_{t,i}$ when the external environment urgency is $\phi_t$;
    \item{\textbf{Transmission Cost $C_2(P_{t,i})$}}. It reflects the cost of transmission power;
     \item{\textbf{Actuation Cost $C_3(C_{t,i})$}}. It quantifies the resources consumed by actuation $C_{t,i}$.
\end{itemize}
Then, the GoT for the $i$-th source can be expressed as follows:
\begin{equation}
GoT_{t,i}^{\pi }=C_1\left( X_{t,i} ,\phi_t \right) +\alpha C_2\left(P_{t,i} \right) +\beta C_3\left( C_{t,i} \right),
\end{equation}
where $\alpha,\beta$ define weighting coefficients for transmission cost and actuation cost, respectively. 

Our goal is to obtain the optimal joint NOMA transmission-control policy, such that the  system’s \textit{long-term average} GoT is minimized. Specifically, the optimization problem can be formulated as follows:
\begin{align}\label{eq:5}
\underset{\pi}{\min}\underset{T\rightarrow \infty}{\lim}\frac{1}{T}\sum_{t=1}^T{\sum_{i=1}^N{\mathbb{E}\left[GoT_{t,i}^{\pi} \right]}},
\end{align}
where $\pi:\mathcal{X}\times\mathcal{T}\mapsto \mathcal{E}\times\mathcal{C}$ represents a joint transmission-control policy determined by the controller, $\mathcal{E}$ denotes transmission power space.

We remark that the remote controller does not have direct access to the controlled source, which renders a typical MDP framework insufficient and poses challenges in effective decision-making. To address this challenge, in what follows we formulate the problem as a \textit{Partially Observable Markov Decision Process} (POMDP) \cite{krishnamurthy2016partially}.

\subsection{POMDP Formulation}\label{sectionIII}
The POMDP is typically characterized by a tuple $\left\{ \mathcal{S} ,\mathcal{A} ,\mathcal{O}, \mathcal{P} , O, r \right\} $. In this subsection, we define each component of this tuple in the context of our problem, detailing how they relate to the controller’s decision-making process under partial observability. 

\begin{itemize}
	\item \textbf{States $\mathcal{S}$.} 
The state of closed-loop $i$ at time slot $t$ is denoted by $\mathbf{s}_{t,i}\triangleq 
\langle X_{t,i},\hat{X}_{t,i},\varDelta_{t,i}\rangle$, which indicates that the global state includes the state of the source, the state received by the controller, and the corresponding age.
 Then, the system state at time slot $t$ can be denoted by $\mathbf{s}_t\triangleq\langle \mathbf{X}_{t},\mathbf{\hat{X}}_{t},\boldsymbol{\varDelta}_{t}\rangle$, where $\mathbf{X}_t\triangleq[X_{t,1},X_{t,2},\cdots,X_{t,N}]$,$\mathbf{\hat{X}}_t\triangleq[\hat{X}_{t,1},\hat{X}_{t,2}\cdots,\hat{X}_{t,N}]$, and $\boldsymbol{\varDelta}_{t}\triangleq[\varDelta_{t,1},\varDelta_{t,2},\cdots,\varDelta_{t,N}]$, respectively.
    \item \textbf{Actions $\mathcal{A}$.} The action of the remote controller at time slot $t$ is denoted by $\mathbf{a}_t\triangleq\langle\mathbf{P}_t,\mathbf{C}_t\rangle$, which consists of two components. $i)$ Power allocation vector $\mathbf{P}_t\triangleq[P_{t,1},P_{t,2},\cdots,P_{t,N}]$, denotes the transmission power for each sensor at time slot $t$. In this study, we discretize the power levels while accounting for the sensors' limited transmission power, giving: $P_{t,i}\in\{0,\frac{P_{\max}}{M},\frac{2P_{\max}}{M},\cdots,1\}$, where $P_{\max}$ represents the sensor's maximum power and $M$ denotes the power level. $ii)$ Actuation vector $\mathbf{C}_{t}\triangleq[C_{t,1},C_{t,2},\cdots,C_{t,N}]$ denotes the commands sent to each actuator at time slot $t$.

     \item\textbf{{Observations $\mathcal{O}$}.} The observation of the controller at time slot $t$ is denoted by $\mathbf{o}_t\triangleq\langle\mathbf{\hat{X}}_t,\boldsymbol{\varDelta}_t\rangle$.

     \item\textbf{{Transition Probabilities {$\mathcal{P}$}}.} The transition probability is defined as $\Pr(\mathbf{s}_{t+1}|\mathbf{s}_t,\mathbf{a}_t)$ representing the conditional probability of the state $\mathbf{s}_{t+1}$ given current state $\mathbf{s}_t$ and action $\mathbf{a}_t$. We remark that the transitions of  $\mathbf{X}_t,\mathbf{\hat{X}}_t$ and $\boldsymbol{\varDelta}_t$ are conditionally independent of each other. Furthermore, their transition probabilities are independent of certain parts of $(\mathbf{s}_t,\mathbf{a}_t)$. Thus, we have:
{\small
\begin{multline}
\Pr(\mathbf{s}_{t+1} | \mathbf{s}_t, \mathbf{a}_t) = 
\Pr(\mathbf{X}_{t+1} | \mathbf{X}_t, \mathbf{C}_t) \cdot \\
\Pr(\mathbf{\hat{X}}_{t+1} | \mathbf{X}_t, \mathbf{\hat{X}}_t,\mathbf{P}_t) \cdot 
\Pr(\boldsymbol{\varDelta}_{t+1} | \boldsymbol{\varDelta}_t, \mathbf{P}_t),
\end{multline}
}

since each closed-loop is independent of the others, we can further obtain:
\begin{equation} \label{eq:6}
\Pr(\mathbf{X}_{t+1} | \mathbf{X}_t, \mathbf{C}_t)=\prod_{i=1}^N{\Pr(X_{t+1,i}|X_{t,i},C_{t,i})},
\end{equation}
{\small
\begin{equation}\label{eq:7}
\Pr(\mathbf{\hat{X}}_{t+1} | \mathbf{X}_t, \mathbf{P}_t)=\prod_{i=1}^N{\Pr(\hat{X}_{t+1,i}|X_{t,i},\hat{X}_{t,i},\mathbf{P}_{t})},
\end{equation}
}and
\begin{equation}\label{eq:8}
\Pr(\boldsymbol{\varDelta}_{t+1} | \boldsymbol{\varDelta}_t, \mathbf{P}_t)=\prod_{i=1}^N{\Pr(\varDelta_{t+1,i}|\varDelta_{t,i},\mathbf{P}_{t})}.
\end{equation}
Each term on the right-hand side of (\ref{eq:6}), (\ref{eq:7}), (\ref{eq:8}) can be obtained by (\ref{eq:1}), 
 (\ref{eq:3}) and (\ref{eq:2}), respectively. Note that the transition probabilities for $\mathbf{\hat{X}}_{t}$ and $\boldsymbol{\varDelta}_t$ require calculating the transmission error rate $\varepsilon_{t,i}$, which depends on power vector $\mathbf{P}_t$, and can be determined using (\ref{eq:10}).

     \item\textbf{{Observation Function $O$}.} The observation function $O$ is defined as $\Pr(\mathbf{o}_t|\mathbf{s}_t,\mathbf{a}_t)$, representing the probability of an observation $\mathbf{o}_t$ given the state $\mathbf{s}_t$ and action $\mathbf{a}_t$. Note that observations at the controller are independent of any actions. Additionally, $\mathbf{\hat{X}}_t$ and $\boldsymbol{\varDelta}_t$, which are related to packets received by the controller, are fully observable. However, the source state $\mathbf{X}_t$ remains completely unobservable to the controller. We then have:
\begin{equation}
\Pr(\mathbf{o}_t|\mathbf{s}_t, \mathbf{a}_t) = 
\begin{cases}
1 & \text{if } \mathbf{o}_t = (\mathbf{\hat{X}}_t,\boldsymbol{\varDelta}_t) \\
0 & \text{otherwise}
\end{cases}
\end{equation}
    \item\textbf{{Reward $r$}.} Our objective is to minimize the system's long-term average GoT. Thus, the reward is defined as the negative instantaneous weighted sum of GoT at time slot $t$, expressed as $r_t\triangleq -\sum\limits_{i=1}^N{GoT_{t,i}}$.
\end{itemize}
\subsection{Belief State}
We note that due to the delayed observed states $\mathbf{\hat{X}}_t=(\hat{X}_{t,1},\cdots,\hat{X}_{t,N})$ at the controller at time slot $t$, the formulated POMDP problem cannot be solved efficiently by similarly applying existing freshness-oriented reinforcement learning-based scheduling frameworks, which are designed for scenarios with delay-free perfect observations (e.g., \cite{bai2023aoi,tang2024learn,10723104}). To tackle this challenge, in what follows we will leverage the posterior probability
distribution of $\mathbf{X}_t$ given the \textit{observation history} $\mathcal{H}_{o,t}$ and the action history $\mathcal{H}_{a,t}$\footnote{The posterior probability
distribution of $\mathbf{X}_t$ is generally a \textit{sufficient statistics} of the \textit{history} $\mathcal{H}_t=(\mathcal{H}_{o,t},\mathcal{H}_{a,t})$, thus it is sufficient to make decisions based on the posterior probability
distribution.}. This posterior probability
distribution is also called as the \textit{belief state} of the POMDP\cite[Chapter 7]{krishnamurthy2016partially}, which preserves the Markov property. In our context, the \textit{belief state} can be derived by calculating the transition probabilities, expressed as:
\begin{equation}\label{eq:11}
    \varpi_{i}^{C_{t-\varDelta_{t}:t-1,i}} = \left( \prod_{l=t-\varDelta_{t}}^{t} \mathbf{P}_i^{C_l} \right)^{\mathbf{T}} \mathbf{u}_{\hat{x}_{t,i}},
\end{equation}
 where $\mathbf{u}_{x}$ represents a one-hot column vector with all elements set to zero except for the element at position $x$, which is set to 1. $\mathbf{P}_i^{C_l}$ denotes the transition matrix for $X_{t,i}$ under actuation $C_l$.
By enabling estimation of the source state, this approach effectively mitigates the challenge of unobservability and supports the development of an optimal joint transmission-control strategy.

\section{Deep Reinforcement Learning Method}\label{section4}
In remote control scenarios, optimal scheduling strategies are crucial for managing an increasing number of source nodes and semantic states. However, as the number of monitored sources grows, the observation space $\mathcal{O}$ expands exponentially, complicating the design of an optimal policy. This complexity is further exacerbated by the lack of perfect channel state information (CSI) in dynamic communication environments, making accurate modeling of transition functions challenging. These issues render traditional methods, such as dynamic programming \cite{bertsekas2012dynamic}, less effective. To overcome these challenges, we adopt a reinforcement learning (RL) approach in this work.
Given our discrete action space, we implement the Double-Dueling Deep Q-Network (D3QN), an advanced variant of DQN. D3QN improves upon DQN by addressing Q-value overestimation through two separate Q-networks, decoupling action selection from value evaluation, which reduces bias in Q-value estimates. Its dueling architecture further refines the Q-function by separating the state-value and advantage functions, leading to more effective evaluations of state values and action advantages. D3QN also retains DQN’s use of experience replay to store past transitions, enabling random mini-batch sampling that reduces sample correlations for improved stability. Moreover, D3QN employs a target network for TD target computation, updated periodically to align with the main network. These enhancements significantly boost stability and performance, making D3QN well-suited for our discrete action space.

\section{Simulation Results}\label{sectionVI}
In this section, we consider 2 controlled monitored source with $N=2$. The size of the source semantic space $\mathcal{X}$, the actuation space $|\mathcal{C}|$, and source context space are set as 2. In the GoT, the status cost function is set as $C_1(x_1,v_1)=C_1(x_1,v_1)=0$, $C_1(x_2,v_1)=1$, $C_1(x_2,v_2)=3$, the transmission cost function is set as $C_2(P_{t,i})=P_{t,i}$, and the actuation cost function is set to be $C_3(c_1)=0$, $C_3(c_2)=1$. The D3QN training parameters and transmission parameters for NOMA uplink are listed in Table \ref{table.10}.
\renewcommand{\arraystretch}{1.2} 
\begin{table}[h]
\centering
\caption{Parameter Settings}\label{table.10}
\begin{tabular}{@{\hspace{15pt}}l@{\hspace{15pt}}c@{\hspace{15pt}}c@{\hspace{15pt}}}
\toprule
\textbf{Parameter}       & \textbf{Symbol} & \textbf{Value}     \\
\midrule
Distance                 & \(d_1,d_2\)     & 2, 5               \\
Path Loss Exponent       & \(\tau\)        & 2                   \\
Message Length           & \(n\)           & 160                 \\
Transmission Symbol      & \(m\)           & 200                 \\
Power Level            & \(M\)           & 20                  \\
\midrule
\vspace{2pt} 
Transition Probabilities of \(X\) & \(p_{x_1,x_2}^{c1}, p_{x_2,x_1}^{c1}\) & 0.05, 0 \\
                                 & \(p_{x_1,x_2}^{c2}, p_{x_2,x_1}^{c2}\) & 0.05, 0.5 \\           
\midrule
Transition Probabilities of \(\phi\) & \(p_{v_1,v_2}, p_{v_2,v_1}\) & 0.5, 0.5 \\
Learning Rate            & \(\lambda\)      & \(10^{-3}\)               \\
Batch Size               & \(|\mathcal{B}|\)           & 64                 \\
Training Episodes                  & \(E\)           & 300                \\
Discount Factor               & \(\rho\)           & 0.95                \\
Replay Buffer Capacity            & \(|\mathcal{D}|\)     & \(3000\) \\
Exploration Rate        & \(\epsilon\)           & 0.4    \\
Synchronous Frequency     & \(k\)           & 100    \\
\bottomrule
\end{tabular}
\end{table}

\begin{figure}
    \centering
   \includegraphics[angle=0,width=0.45\textwidth]{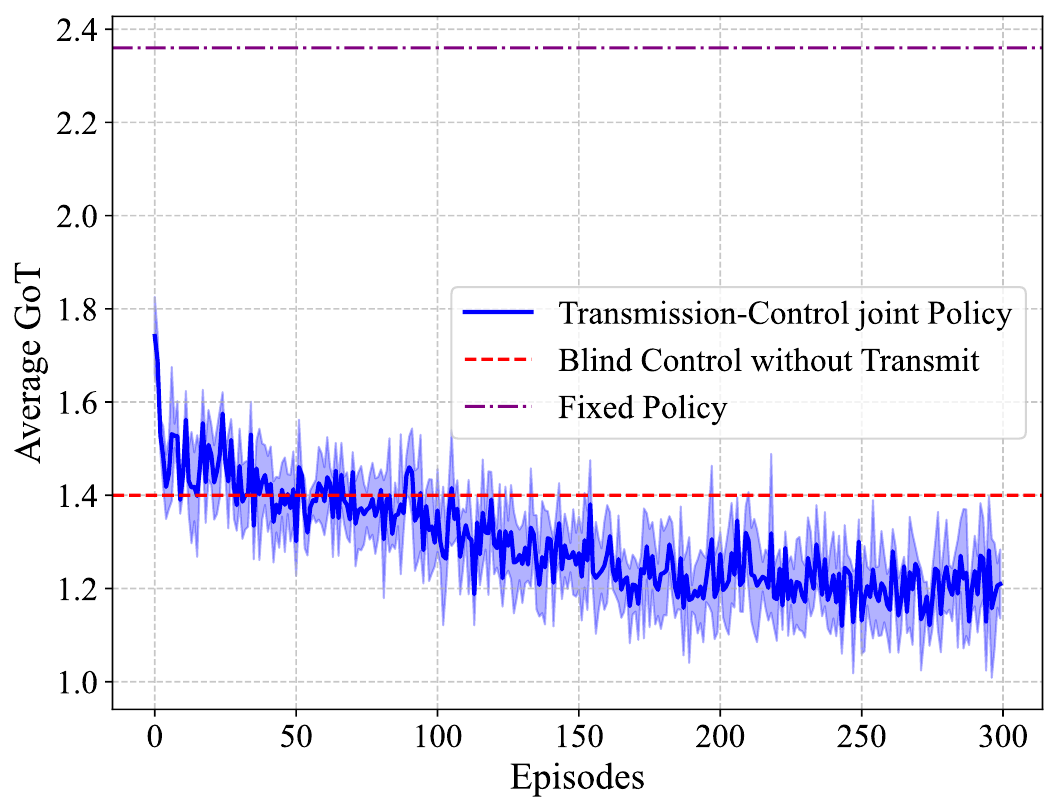}
    \caption{Learning curve of D3QN and comparison of baselines when SNR = 17dB, $\alpha=0.2$.}
    \label{fig:3}
\end{figure}

 Fig. \ref{fig:3} illustrates the training results of D3QN algorithm. We conducted 10 training iterations, where the solid line denotes the average performance across these trials, and the shaded region illustrates the training variance. It is evident that the average GoT exhibits a declining trend as training advances. Following 200 episodes, the GoT steadily converges as the agent’s exploration rate diminishes.

 To validate the transmission-control joint policy, we compare it with two benchmarks: 
 \(a\)) \textit{Fixed Policy}: the system does not transmit and uses a fixed control strategy \(c_2\); \(b\)) \textit{Blind Control without Transmit}: the system refrains from transmission and optimizes the control strategy using reinforcement learning based on source state estimates as in (\ref{eq:11}), representing a scenario with poor communication links. The results show that the transmission-control joint strategy, trained with the D3QN algorithm, significantly outperforms both baselines.


\begin{figure}
    \centering
    \includegraphics[angle=0,width=0.45\textwidth]{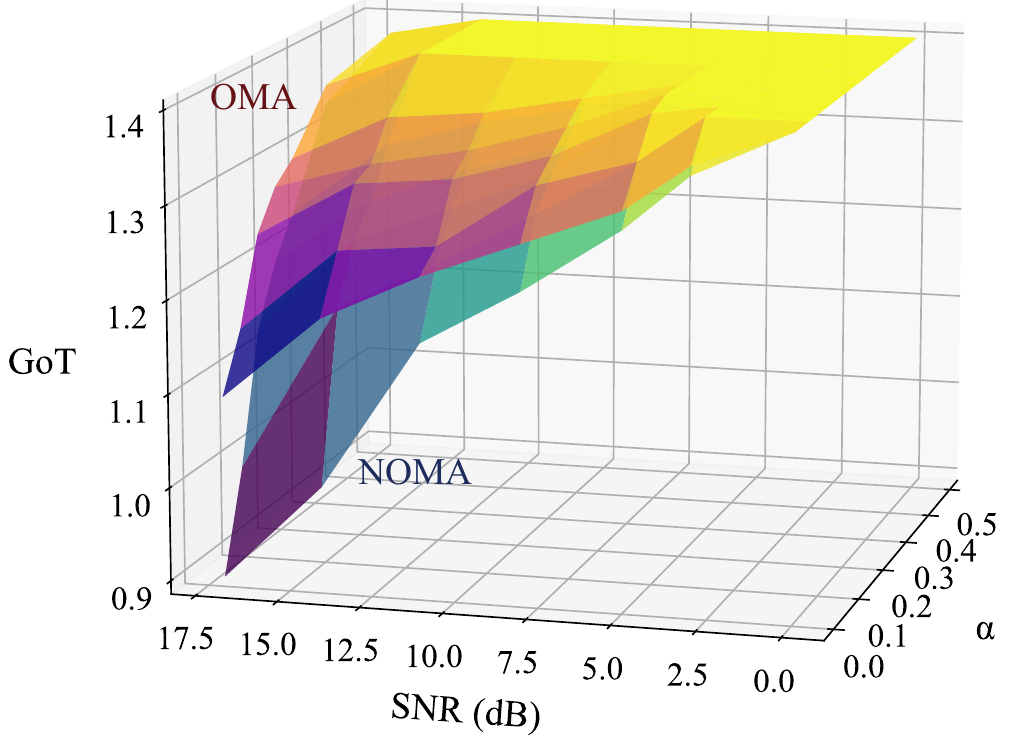}
    \caption{The GoT performance comparison of NOMA and OMA under different SNR and transmission cost coefficients $\alpha$.}
    \label{fig:5}
\end{figure}

To investigate the impact of different access methods for transmission on overall system utility, we compared the GoT performance of NOMA and OMA under various SNR conditions and different transmission cost coefficients $\alpha$. As shown in Fig. \ref{fig:5}, NOMA consistently outperforms OMA in GoT performance across all channel conditions and various transmission cost coefficients. This advantage arises from NOMA's capability to transmit multiple state packets simultaneously, while OMA is restricted to transmitting only one state packet. Consequently, in a NOMA system, the controller can receive timely updates from multiple corresponding sources, facilitating a more accurate estimation of multi-source states and enhancing the precision of control for each loop.

Specifically, at high SNR and low $\alpha$, NOMA substantially outperforms OMA. However, as SNR decreases and $\alpha$ increases, the advantages of NOMA gradually diminish, eventually aligning with those of OMA. This is because, under favorable channel conditions and low transmission costs, the system is more inclined to expend power for successful transmission, thereby enhancing control performance. In this context, NOMA's advantage of simultaneously transmitting multiple source states becomes particularly evident. Conversely, when channel conditions are poor and transmission costs are high, the control gains from transmission are outweighed by the power costs, leading the system to reduce the resources allocated for transmission. In such cases, the reliability of NOMA is compromised, leading to a diminished performance advantage.

\begin{figure}
    \centering
    \includegraphics[angle=0,width=0.45\textwidth]{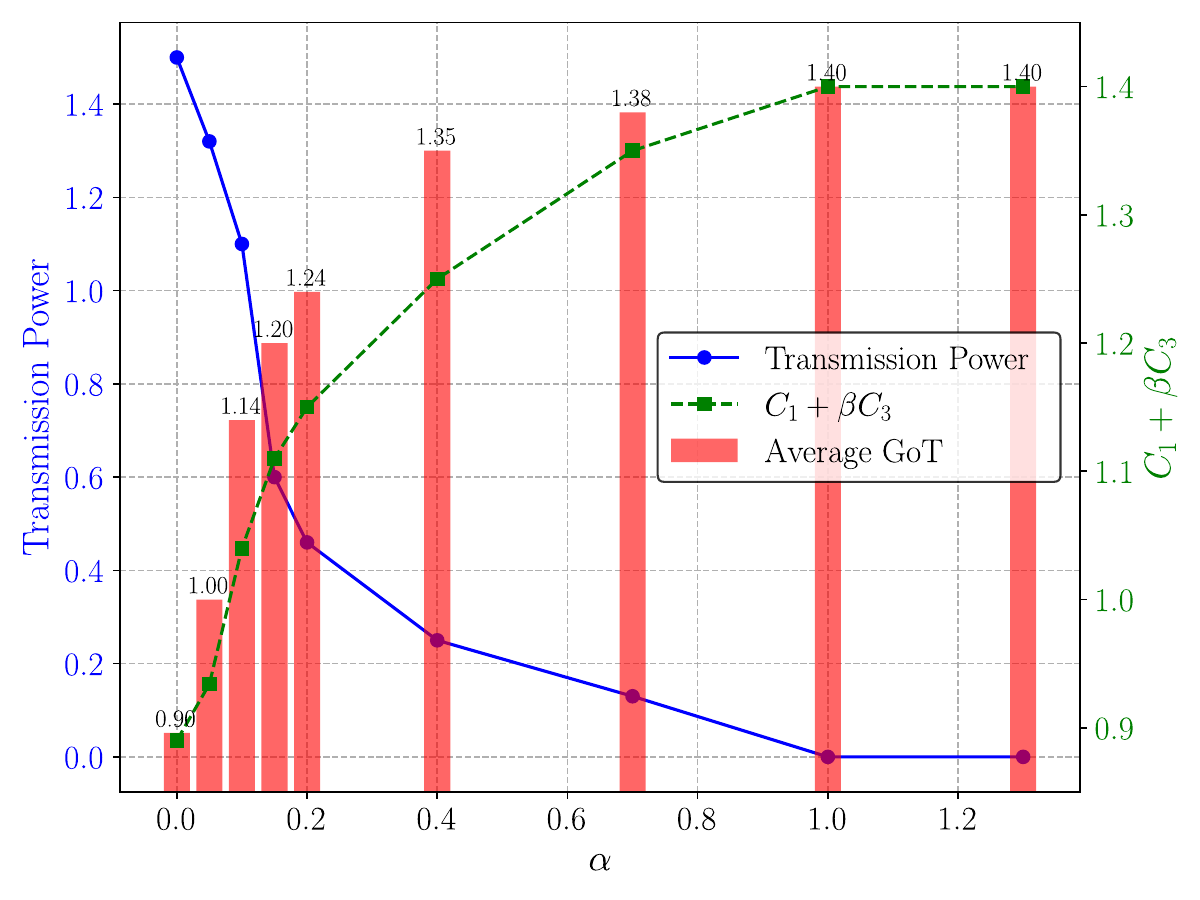}
    \caption{The transmission power, $C_1+\beta C_3$, and average GoT under different transmission cost coefficient $\alpha$ with SNR = 17dB, where $C_1+\beta C_3$ evaluats the effectiveness of the system's control performance.}
    \label{fig:4}
\end{figure}

Fig. \ref{fig:4} illustrates the trade-off between transmission and control in a remote control closed-loop system. \(C_1\) represents the state cost, reflecting the benefits of control, while \(C_3\) indicates the inherent costs of implementing control commands. Thus, \(C_1 + \beta C_3\) evaluates the system's control performance. As the transmission cost coefficient \(\alpha\) increases, the system reduces its power allocation for transmission, causing the controller to receive less frequent state updates. This leads to inaccurate inferences of the source's status, hindering the generation of optimal control commands. Consequently, the control cost rises, signaling a decline in system performance. This demonstrates the crucial role of transmission in remote control systems and the inherent trade-off between transmission efficiency and control effectiveness.
 Moreover, the system's GoT continues to increase with rising transmission cost coefficients $\alpha$ until it attains a maximum threshold. Beyond this point, due to the excessive transmission costs, the system ceases all transmission (zero transmission power consumption), and the controller resorts to blind control.

\section{Conclusion}
In this work, we have accomplished a joint optimization of transmission and control within a Goal-oriented NOMA network. We have introduced a novel metric, the Goal-oriented Tensor (GoT), to effectively quantify system utility in remote control scenarios. By formulating the GoT-optimal problem as a POMDP and leveraging a D3QN algorithm, our approach has achieved intelligent power allocation and optimized control decisions. We further reveal a fundamental trade-off between transmission efficiency and control fidelity in remote systems, with NOMA demonstrating distinct advantages over OMA in enhancing remote control performance. These contributions highlight NOMA’s promise in advancing utility for multi-user remote control systems, establishing a foundation for further exploration in goal-oriented network design.

\end{document}